\documentclass[aps,prl,twocolumn,amsmath,amssymb,superscriptaddress,floatfix]{revtex4-1}
\usepackage{amsmath}
\usepackage{amssymb}
\usepackage{amsfonts}
\usepackage{listings}
\usepackage{enumerate}
\usepackage{latexsym}
\usepackage{bm}
\usepackage{graphicx}
\usepackage{braket}
\usepackage[pdftex,colorlinks=false]{hyperref}
\usepackage{xcolor}
\usepackage{verbatim}
\usepackage{float}

\usepackage{times}

\newcommand{\beq}{\begin{equation}}
\newcommand{\eneq}{\end{equation}}
\newcommand{\bs}[1]{\boldsymbol{#1}}

\def\be{\begin{equation}}
\def\ee{\end{equation}}
\def\ba{\begin{eqnarray}}
\def\ea{\end{eqnarray}}

\def\R{{\rm Re}}
\def\Z{\mathbb{Z}}
\def\C{\mathbb{C}}

\def\beq{\begin{equation}}
\def\eeq{\end{equation}}
\def\barray{\begin{eqnarray}}
\def\earray{\end{eqnarray}}

\font\upright=cmu10 scaled\magstep1
\def\stroke{\vrule height8pt width0.4pt depth-0.1pt}

\def\Zmath{\mathbb{Z}}
\def\Qmath{\vcenter{\hbox{\upright\rlap{\rlap{Q}\kern
                   3.8pt\stroke}\phantom{Q}}}}
\def\Nmath{\vcenter{\hbox{\upright\rlap{I}\kern 1.7pt N}}}
\def\Cmath{\vcenter{\hbox{\upright\rlap{\rlap{C}\kern
                   3.8pt\stroke}\phantom{C}}}}
\def\Rmath{\vcenter{\hbox{\upright\rlap{I}\kern 1.7pt R}}}
\def\Z{\ifmmode\Zmath\else$\Zmath$\fi}
\def\Q{\ifmmode\Qmath\else$\Qmath$\fi}
\def\N{\ifmmode\Nmath\else$\Nmath$\fi}
\def\C{\ifmmode\Cmath\else$\Cmath$\fi}
\def\R{\ifmmode\Rmath\else$\Rmath$\fi}

\input{epsf}

\newcounter{defcounter}
\begin{document}

\tolerance 10000

\newcommand{\cbl}[1]{\color{blue} #1 \color{black}}

\newcommand{\vk}{{\bf k}}

\widowpenalty10000
\clubpenalty10000

\title{Symmetries and many-body excited states with neural-network quantum states}

\author{
Kenny~Choo}
\address{
 Department of Physics, University of Zurich, Winterthurerstrasse 190, 8057 Zurich, Switzerland
}

\author{Giuseppe Carleo}
\address{
Center for Computational Quantum Physics, Flatiron Institute, 162 5th Avenue, New York, NY 10010, USA
}

\author{Nicolas Regnault}
\address{
Laboratoire Pierre Aigrain, Ecole normale sup\'erieure, PSL University, 
Sorbonne Universit\'e, Universit\'e Paris Diderot, Sorbonne Paris Cit\'e, 
CNRS, 24 rue Lhomond, 75005 Paris France
}

\author{
Titus~Neupert}
\address{
 Department of Physics, University of Zurich, Winterthurerstrasse 190, 8057 Zurich, Switzerland
}

\begin{abstract}
Artificial neural networks have been recently introduced as a general ansatz to compactly represent many-body wave functions. 
In conjunction with Variational Monte Carlo, this ansatz has been applied to find Hamiltonian ground states and their energies. Here we provide extensions of this method to study properties of excited states, a central task in several many-body quantum calculations. First, we give a prescription that allows to target eigenstates of a (nonlocal) symmetry of the Hamiltonian. Second, we give an algorithm that allows to compute low-lying excited states without symmetries. We demonstrate our approach with both Restricted Boltzmann machines states and feedforward neural networks as variational wave-functions. Results are shown for the one-dimensional spin-$1/2$ Heisenberg model, and for the one-dimensional Bose-Hubbard model.  When comparing to available exact results, we obtain good agreement for a large range of excited-states energies. Interestingly, we also find that deep networks typically outperform shallow architectures for high-energy states.  \end{abstract}

\date{\today}

\maketitle

\textit{Introduction} ---
Artificial neural networks (ANN) for machine learning (ML) are quickly becoming an indispensable tool not just in every-day life applications like voice recognition, but also in fundamental sciences. In the context of applied statistical physics, for instance, machine learning techniques have been used successfully for classifying phases of matter and phase transitions \cite{Nieuwenburg2017,Carrasquilla2017,Schindler2017,YoshiokaML2018,KaubrueggerBudich2018,KimML2018}, speeding up Monte Carlo simulations \cite{ShenLiuFu2018,HuangWangMC2017}, molecular modelling \cite{RuppAlexandreML2012,de_comparing_2016} and more. These applications are close in spirit to classical ML tasks, in that the networks are trained using labelled data such that they learn to approximate a certain target function known on a finite number of data points. 
In the context of many-body quantum physics, a representation of the many-body wave-function based on ANN has been proposed in Ref.~\cite{Carleo2017}. ANN representations can be used in unsupervised applications of ANN and ML, where no labelled data is given a-priori. Applications in this sense include the simulation of ground-states ~\cite{Carleo2017,Saito2017,NomuraMasatoshi2017,GlasserCiracNN2018,DengLiDasSarma2017,SaitoKato2018,carleo_constructing_2018}, and the reconstruction of quantum states from experimental measurements \cite{TorlaiCarleoTomo2018, rocchetto_learning_2018}.

The key difficulty in many-body problems is the exponential growth of the Hilbert space dimension with the system size, leading to an exponential number of parameters needed for an exact representation of the wavefunction. This exponential growth can be, to some extent, circumvented in interesting physical applications using either stochastic sampling approaches, or compact representations of the many-body states. Popular approaches belonging to the two categories are, respectively, Quantum Monte Carlo methods \cite{ceperley_quantum_1986,becca_quantum_2017}, and tensor-network approaches \cite{White2005,Verstraete2008}. Known limitations of these approaches are however the sign problem~\cite{Troyer2005} for Quantum Monte Carlo, and the entanglement problem for tensor networks. As a result, several interesting systems and physical regimes are currently inaccessible by state of the art techniques, including key strongly-interacting fermionic problems in two dimensions, out-of-equilibrium-dynamics, and excited states. 
The learning scheme proposed in Ref.~\onlinecite{Carleo2017} leverages the ability of ANN to compactly represent highly dimensional functions, thus belongs to the second category of variational wavefunction approaches. A distinct feature of this approach is its ability to capture longer range correlations and entanglement structures~\cite{DengLiDasSarma2017} leading to highly accurate representations many-body states \cite{Carleo2017,TorlaiCarleoTomo2018,NomuraMasatoshi2017,ChiralTopoNN2018,CaiLiu2018,GlasserCiracNN2018}.

Previous works~\cite{Carleo2017,Saito2017,NomuraMasatoshi2017,GlasserCiracNN2018,DengLiDasSarma2017,SaitoKato2018} mainly focused on obtaining ground states with ANN variational quantum states. However, for the method to become a comprehensive tool for quantum many-body calculations, it is crucial to have controlled access to -- at least -- low-lying excited states. This is needed to answer questions such as: Is the ground state gapped or gapless? What is the ground state degeneracy? What are the structure and the dispersion of low-lying excitations? 
In this paper, we use ANN variational quantum states to compute excited states and target states with fixed quantum numbers (e.g., with a certain momentum). We achieve this in two ways, first by taking advantage of Abelian spatial symmetries such as translational symmetry and second by orthogonalizing the wavefunction with respect to the ground state. We demonstrate our approach with both Restricted Boltzmann Machine (RBM) states and 3-layer feedforward neural networks (FFNN) as variational wave-functions. 
We test the methods on the one-dimensional spin-$1/2$ Heisenberg model with up to $L=36$ sites and on the one-dimensional Bose-Hubbard model with up to $L=40$ sites at filling one. When comparing to available exact results, we obtain relative errors of about $10^{-5}$--$10^{-3}$ on the variational energies. 

\textit{Restricted Boltzmann Machine} ---
For concreteness, consider a system made of $L$ spin-1/2 degrees of freedom denoted by $\sigma_j=\pm 1$, $j=1,\cdots, L$.
RBMs were proposed in Ref.~\onlinecite{Carleo2017} as a variational ansatz for the many-body wave function of such a system.
The value of the RBM wave function can be represented by a sum of exponentials,
\begin{equation}
\begin{split}
\Psi(\boldsymbol{\sigma}) &= \sum_{\boldsymbol{h}} e^{\sum_{j} a_{j}\sigma_{j} + \sum_{i} b_{i}h_{i} + \sum_{ij} h_{i}W_{ij}\sigma_{j}},
\end{split}
\label{eq: wave function interpretation}
\end{equation}
where the sum runs over all $\bs{h}=(h_1, h_2, \dots h_M)$ with the binary variables $h_i \in\lbrace -1, 1 \rbrace$ for $i=1,\dots,M$.
The neural network in Eq.~\eqref{eq: wave function interpretation} may admit a statistical physics interpretation. The physical spins $\boldsymbol{\sigma}$ are called a visible layer, and $\bs{h}$ is interpreted as a second -- hidden -- layer of auxiliary spins.
The visible and hidden layers can then be considered to interact through an Ising type interaction,
\begin{equation}
E(\boldsymbol{\sigma}, \boldsymbol{h}) = -\sum_{j} a_{j}\sigma_{j} - \sum_{i} b_{i}h_{i} - \sum_{ij} h_{i}W_{ij}\sigma_{j},
\label{eq: energy}
\end{equation}
where $a_{j}$ and $b_{i}$ are known as the visible and hidden bias, respectively, analogous to a local magnetic field, and $W_{ij}$ are the weights corresponding to interactions between visible and hidden nodes.
Equation~\eqref{eq: energy} has the interpretation of a classical energy if the network parameters are taken to be real-valued. Here, to apply the formalism to general wave functions, complex-valued weights and biases are used \cite{Carleo2017}. In this case, Eq.~\eqref{eq: energy} does not have an analogue in classical statistical physics. 

Upon performing the summation over the hidden variables $\boldsymbol{h}$, Eq.~\eqref{eq: wave function interpretation} reduces to
\begin{equation} \label{eq: log rbm}
\log\Psi(\boldsymbol{\sigma}) = \sum_{j} a_{j}\sigma_{j} + \sum_{i} \log \left[\cosh\left( b_{i} + \sum_{j} W_{ij}\sigma_{j} \right) \right]
\end{equation}
up to some additive constant which corresponding to an overall normalization and phase factor of the wave-function.

\textit{Feedforward Neural Network} ---
The second type of network that we consider is a FFNN. The input to the network is a configuration $\boldsymbol{\sigma}$, indexing the many-body basis states. It could be a binary vector for a spin half system or a vector of integers for spinless bosons. 

We construct an $\ell$-layer FFNN as follows. Let $\boldsymbol{v}_{n}$ be the $M_n$-component vector output from layer $n$ and define $\boldsymbol{v}_{0} = \boldsymbol{\sigma}$, where $M_n$ is the number of neurons in layer $n$. At each layer, we apply an affine map followed by element-wise nonlinear function $f$ (the so-called activation function)
\begin{equation}\label{feedforward}
\boldsymbol{v}_{n} \rightarrow \boldsymbol{v}_{n+1} =f\left(\boldsymbol{W}_{n}\boldsymbol{v}_{n} + \boldsymbol{b}_{n}\right),
\end{equation}
where $\boldsymbol{W}_{n}$ is a matrix of size $M_{n+1}\times M_{n}$ known as the weight matrix and $\boldsymbol{b}_{n}$ is a vector called the bias. The activation function $f$ can be chosen freely. Since we would like the ansatz to be generic, we again have to allow for complex valued parameters. Inspired by the effectiveness of the RBM, we choose
$
f(x) = \log\left[\cosh(x)\right]
$ from here on.

The final layer consists of only one neuron so the output $\boldsymbol{v}_{\ell}$ is a one-dimensional vector which corresponds to the value $\boldsymbol{v}_{\ell} = \textrm{log}\left[\Psi(\boldsymbol{\sigma})\right]$. For the case of a single hidden layer followed by the final output layer, the ansatz can be written as
\begin{equation}
\log{\Psi(\boldsymbol{\sigma})} = \boldsymbol{W}_{1}[f( \boldsymbol{W}_{0}\boldsymbol{\sigma} + \boldsymbol{b}_{0})].
\end{equation}
which reduces to the RBM in Eq.~\eqref{eq: log rbm} without visible bias, if  $\boldsymbol{W}_1=(1,1,1,\cdots)$. We do observe in our tests that the single hidden layer FFNN has a similar performance to the RBM. Therefore, to go beyond the RBM we focus on a FFNN with 3 layers: 2 hidden layers followed by an output layer. The ansatz becomes
\begin{equation} \label{eq: FFNN}
\log{\Psi(\boldsymbol{\sigma})} = \boldsymbol{W}_{2}[f( \boldsymbol{W}_{1}[f(\boldsymbol{W}_{0}\boldsymbol{\sigma} + \boldsymbol{b}_{0})]+ \boldsymbol{b}_{1})].
\end{equation}

\textit{Abelian Symmetries} ---
We now explain how to enforce that the network represents an eigenstate of a symmetry of the system.
Let $\lbrace \hat{T}_{1}, \cdots, \hat{T}_{\nu} \rbrace$ be the generators of a finite Abelian symmetry group $G$ of order $\nu$, where the elements $g \in G$ act on the configurations of the system as $g \boldsymbol{\sigma} = \boldsymbol{\sigma}'$. Since $G$ is Abelian, its irreducible representations are purely one-dimensional. A wavefunction belongs to an irreducible representation with character $\lbrace \omega_{1}, \cdots, \omega_{\nu} \rbrace$ corresponding to the $\nu$ generators if
\begin{equation}\label{eq: Irrep}
\begin{split}
 \hat{T}_{i} \ket{\Psi} = \omega_{i} \ket{\Psi} 
\quad
\implies &\quad \bra{\boldsymbol{\sigma}} \hat{T}_{i} \ket{\Psi} = \omega_{i} \braket{\boldsymbol{\sigma}|\Psi} \\
\implies &\quad \Psi(\hat{T}_{i}\boldsymbol{\sigma}) = \omega_{i} \Psi(\boldsymbol{\sigma}). \\
\end{split}
\end{equation}

\begin{figure*}[t]
            \includegraphics[width=1.0\textwidth]{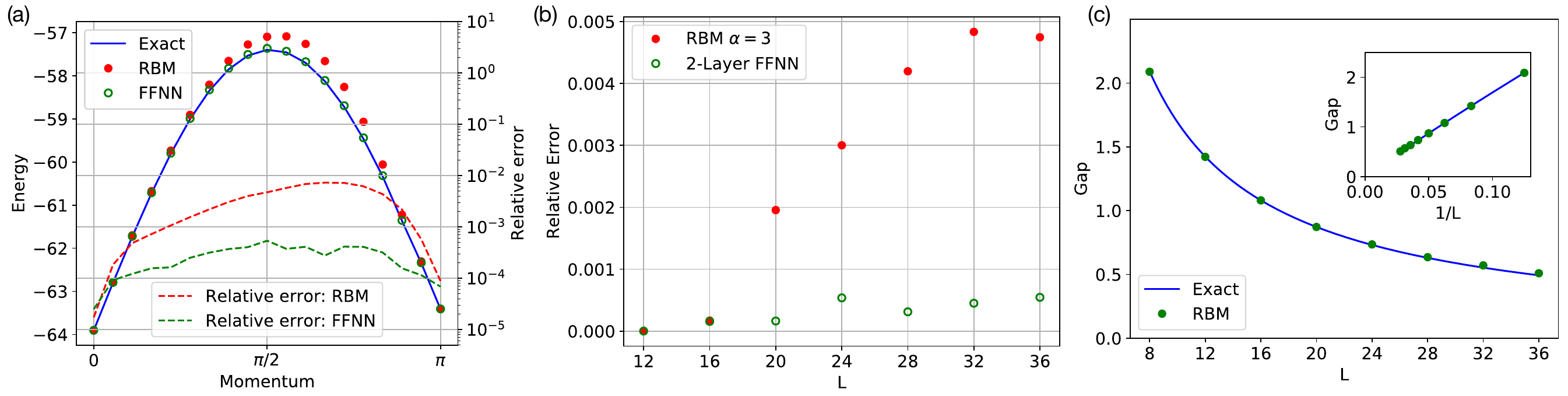}
    \caption{ (a) Momentum-resolved  spectrum of the one-dimensional Heisenberg model with $L=36$ spins. The blue line shows the exact values computed using ED, the green circles represent the energy obtained from a 3-layer FFNN with hidden unit density $\alpha_1=2$ ($\alpha_2=0.5$)  in the first (second) hidden layer (corresponding to $3996$ free parameters) and red dots shows the energy from an RBM with hidden unit density $\alpha_1=3$. (b) Relative error $\epsilon$ as a function of system size, for the $k=\pi/2$ state. For the RBM, we fix the hidden unit density $\alpha_1 =3$, whereas for the FFNN we use a density of $\alpha_1 = 2$ in the first hidden layer and a density of $\alpha_2 = 0.5$ in the second hidden layer.  For the $k=0$ sector the relative error is $\sim 10^{-5}$. (c) Energy gap from the ground state to the first excited state of one-dimensional spin-$1/2$ Heisenberg model. The blue line shows the exact values computed using ED, the green circles represents the energy gap obtained from an RBM with hidden unit density $\alpha = 2$. FFNN results are identical to the RBM ones and are thus not shown here. In the inset, we plot versus $1/L$, showing that the gap is inversely proportional to system size. The relative error of the excited states obtained is less than $3\times10^{-4}$ for all cases.  
    }
\label{fig: Heisenberg Spectrum}
\end{figure*}

In order to obtain the eigenstate corresponding to this irreducible representation, we want that the networks output obeys Eq.~\eqref{eq: Irrep}. Since the network represents the logarithm of the wavefunction, this means $\log{\Psi(\hat{T}_{i}\boldsymbol{\sigma})} = \log\omega_{i} + \log{\Psi(\boldsymbol{\sigma})} $. Due to the highly nonlinear form of the wave function representation, it is not straightforward to adjust the weights of the network such that this condition is strictly satisfied. 
Instead, we solve the problem of obtaining a neural network representation with a specific eigenvalue $\omega_i$ as follows: Let $ \log{\tilde{\Psi}(\boldsymbol{\sigma})}$ represent the value obtained from the network as given by Eq.~\eqref{eq: log rbm} or~\eqref{eq: FFNN}. Next, consider the equivalence classes of configurations related by the symmetry group $G$, i.e., $[\boldsymbol{\sigma}] = \lbrace g\boldsymbol{\sigma} : \forall g \in G\rbrace$. For each equivalence class, we pick a canonical configuration $\boldsymbol{\sigma}_{\mathrm{canonical}}$. We then define the amplitude of a configuration $\boldsymbol{\sigma}$ to be
\begin{equation}
\log{\Psi(\boldsymbol{\sigma})} = \sum_{i=1}^\nu {r_{i,\bs{\sigma}}} \log\omega_{i} + \log\left[\tilde{\Psi}(\boldsymbol{\sigma}_{\mathrm{canonical}})\right],
\label{eq: symm neural network}
\end{equation}
where the integers $r_{i,\bs{\sigma}}$ are the number of times the generator $\hat{T}_i$ needs to be applied to map the canonical configuration back to $\bs{\sigma}$. They are implicitly defined though the equation $ \boldsymbol{\sigma} = \prod_{i=1}^\nu  \hat{T}_{i}^{r_{i,\bs{\sigma}}} \boldsymbol{\sigma}_{\mathrm{canonical}}$.
Such a procedure guarantees that the condition in Eq.~\eqref{eq: Irrep} is satisfied. Then, instead of evaluating $\log{\Psi(\boldsymbol{\sigma})}$ directly for generic $\bs{\sigma}$, we evaluate $\log{\Psi_{N}}(\boldsymbol{\sigma}_{\mathrm{canonical}})$ and obtain $\log{\Psi(\boldsymbol{\sigma})}$ from Eq.~\eqref{eq: symm neural network}. Minimizing the energy with this expression for $\log{\Psi(\boldsymbol{\sigma})}$ in turn gives the lowest eigenstate in the selected symmetry sector.

Let us illustrate this procedure for translational symmetry in one dimension. In this case, there is only a single generator $\hat{T}$. Then for a state $\ket{\Psi}$ with momentum $k$, the amplitude of a configuration $\boldsymbol{\sigma}$ is given by 
\begin{equation}
\log{\Psi(\boldsymbol{\sigma})}  = \mathrm{i}r_{\bs{\sigma}}k +  \log\left[\tilde{\Psi}(\boldsymbol{\sigma}_{\mathrm{canonical}})\right]
\end{equation}
where $\boldsymbol{\sigma} = \hat{T}^{r_{\bs{\sigma}}} \boldsymbol{\sigma}_{\mathrm{canonical}}$. 
~\footnote{We choose the canonical configuration to be the lexicographically smallest one, for example
$
\boldsymbol{\sigma}  = (1,0,1,1,0,0) \rightarrow (0,0,1,0,1,1) = \hat{T}^{2} \boldsymbol{\sigma} = \boldsymbol{\sigma}_{\mathrm{canonical}}
$.
}

\textit{Excited States Without Symmetry} ---
Many physical problems of interest possess (nearly) degenerate ground states that are not distinguished by good quantum numbers, for instance in topologically ordered systems or in spin glasses. In this case, the following procedure can be applied. 
The task is as follows: Given an ANN variational wave function which represents the ground state of a Hamiltonian, say $\Phi_{0}(\boldsymbol{\sigma})$ we would like to find the wavefunction $\Psi$ with the lowest energy but orthogonal to $\Phi_{0}$. To that end, we define
\begin{equation} \label{eq: psi sum}
\Psi = \Phi_{1} - \lambda \Phi_{0}
\end{equation}
where $\lambda$ is a complex scalar and $\Phi_{1}$ corresponds to a different ANN variational wave function with its own set of weights and biases. To enforce orthogonality between $\Psi$ and $\Phi_{0}$, i.e. $\braket{\Phi_{0}|\Psi}=0$ we set 
$
\lambda = \frac{\braket{\Phi_{0}|\Phi_{1}}}{\braket{\Phi_{0}|\Phi_{0}}},
$
which can be computed in standard Monte Carlo fashion
\begin{equation} \label{eq: lambda}
\lambda =  \sum_{\boldsymbol{\sigma}} \left( \frac{\Phi_{1}(\boldsymbol{\sigma})}{\Phi_{0}(\boldsymbol{\sigma})} \right) \frac{|\Phi_{0}(\boldsymbol{\sigma})|^2}{\sum_{\boldsymbol{\sigma'}}|\Phi_{0}(\boldsymbol{\sigma'})|^2} \approx  \Bigg\langle \frac{\Phi_{1}(\boldsymbol{\sigma})}{\Phi_{0}(\boldsymbol{\sigma})} \Bigg\rangle_{N_{\mathrm{s}}}
\end{equation}
where the average is carried over $N_{\mathrm{s}}$ samples generated from the distribution $|\Phi_{0}(\boldsymbol{\sigma})|^2$ through  Monte Carlo sampling. 

The optimization scheme then proceeds in two steps:
\begin{enumerate}
\item Sample the ground state wavefunction $|\Phi_{0}(\boldsymbol{\sigma})|^2$ to compute $\lambda$ as in Eq.~\eqref{eq: lambda}
\item Perform the imaginary time evolution with stochastic reconfiguration \cite{Sorella2007SR} on the full wavefunction $\Psi = \Phi_{1} - \lambda \Phi_{0}$ using the updated $\lambda$.
\end{enumerate}
In principle, if the overlap $\lambda$ can be computed exactly and the stochastic reconfiguration step is not subject to sampling noise or approximation errors [see Eq.~\eqref{eqn: exp}], one only needs to project away the ground state component once and then the imaginary time evolution would necessarily converge to the first excited state. However, due to the various sources of noise and possibly even numerical errors,  any small component of the ground state would grow exponentially, thus making it necessary to constantly perform the projection.

Finally, it is important to note that due to the stochastic nature of the optimization, the state $\Psi$ is not exactly orthogonal to the ground state. In order to quantify the accuracy of the result we can compute the normalized overlap
\begin{equation}
\frac{\braket{\Phi_{0}|\Psi}}{\braket{\Phi_{0}|\Phi_{0}}}  \frac{\braket{\Psi|\Phi_{0}} }{\braket{\Psi|\Psi}},
\end{equation}
which can also be computed as a Monte Carlo average. 

\textit{Results} ---
To test the two methods introduced above, we study two one-dimensional benchmark models: the spin-$1/2$ antiferromagnetic Heisenberg chain and the Bose-Hubbard chain.
The former is defined by the Hamiltonian
\begin{equation}
\hat{H} = 4\sum_{i=1}^{L} \boldsymbol{\hat{S}}_{i}\cdot\boldsymbol{\hat{S}}_{i+1},
\end{equation}
where $\boldsymbol{\hat{S}}_i$ are the spin-$1/2$ operators on site $i$ and we choose periodic boundary conditions. The momentum-resolved spectrum of this model can be obtained using the Bethe ansatz~\cite{KlumperBethe1991}.

\begin{figure*}[t]
            \includegraphics[width=0.8\textwidth]{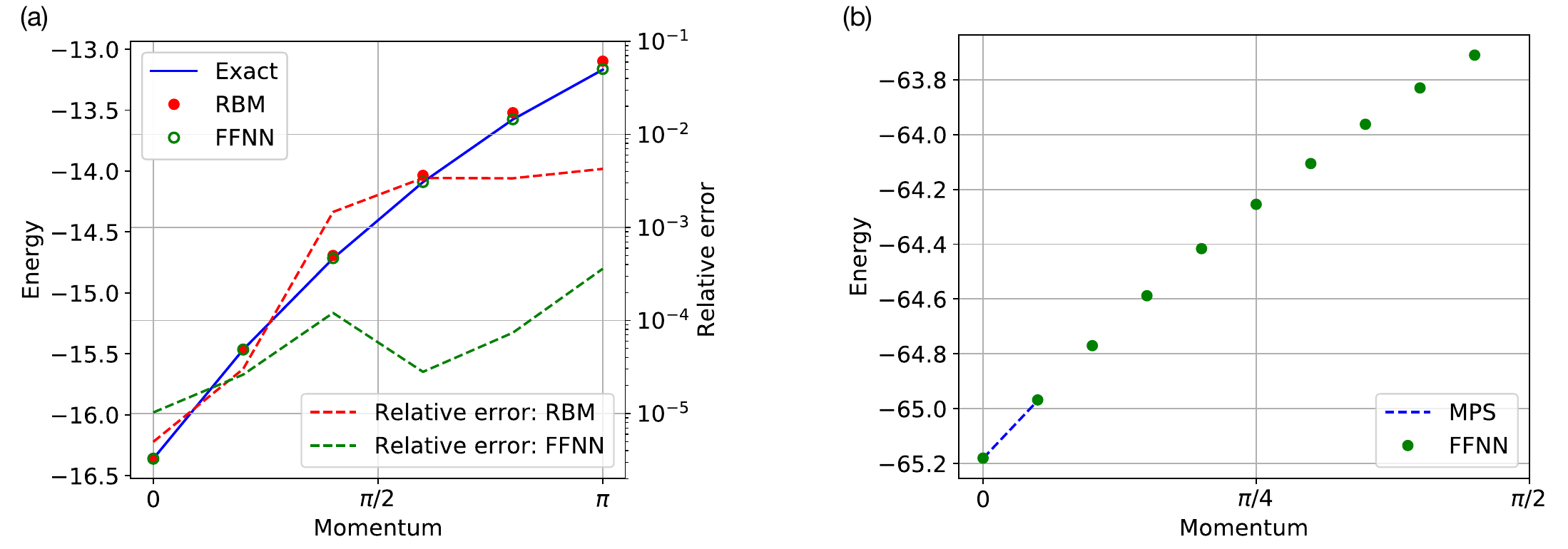}
    \caption{Momentum-resolved  spectrum of weakly interacting $U=1$ bosons on a one-dimensional periodic lattice. (a) $N=10$ bosons in $L=10$ sites. Blue line shows the analytically calculated value and the green circles indicates the value obtained from a 3-layer FFNN with hidden unit density $\alpha_1 = 4$ ($\alpha_2 = 1$) in the first (second) hidden layer ($860$ free parameters). The red circles show the value from an RBM with hidden unit density $\alpha_1 = 8$ ($890$ free parameters).  Dashed lines indicate the relative error. (b) $N=40$ bosons in $L=40$ sites. Dashed blue line shows values inferred from MPS calculations. The green circles indicates the values obtained from a standard fully connected 3-layer feedforward neural network with hidden unit density $\alpha_1=2$ ($\alpha_2 = 1$) in the first (second) hidden layer ($6560$ free parameters) except for the last point $k=\frac{18\pi}{40}$ where we used 100 neurons in the first hidden layer and 40 in the second layer ($8180$ free parameters). We show only the first 10 momenta.}
\label{fig: Hubbard Spectrum}
\end{figure*}

As a first benchmark, we computed the momentum spectrum of the model with $L=36$ sites using both the RBM and the 3-layer deep FFNN and compared them to the results from exact diagonalization (ED). We set the hidden unit density defined by $\alpha_n = M_n/L$ to be $\alpha_1=3$ for the RBM and $\alpha_1=2$ ($\alpha_2=0.5$) for the first (second) layer of the FFNN. The ANN results, compared to those obtained from ED are shown in Fig.~\ref{fig: Heisenberg Spectrum}(a). One can observe that the relative error $\epsilon = \left| (E - E_{\textrm{exact}})/E_{\textrm{ground}} \right|$ is much larger for higher energy states, i.e., for momenta away from $0$ or $\pi$. Moreover, the relative error for the RBM is higher than that of the 3-layer FFNN, possibly suggesting that either the RBM ansatz is less efficient at representing those excited states or that the optimization of the network is caught in a local minimum. We checked that increasing the number of hidden units systematically improves the accuracy of the network. 

In Fig.~\ref{fig: Heisenberg Spectrum}(b), we show the scaling of the relative error with system size for the two different network architectures, which shows that the 3-layer FFNN systematically performs better than an RBM with a comparable number of parameters.  
Whereas the relative error remains roughly constant with system size for the FFNN,  the RBM error instead seems to increase linearly. Once again, this circumstance does not strictly imply that RBM machines are less expressive than FFNNs, since optimization is also an extremely crucial ingredient to be considered.

Next, using the two-step method described above, we obtained the energy gap from the ground state to the first excited state as a function of system size $L$. This way, we do not use any information about the translation symmetry. Exact values were computed using ED. The results are shown in Fig.~\ref{fig: Heisenberg Spectrum}~(c).  Here, the hidden unit density of $\Phi_{1}$ [see Eq.~\eqref{eq: psi sum}] was fixed at $\alpha_1 =2$ (except the $L=40$ computation where we used $\alpha_1 = 4$), while the ground state $\Phi_{0}$ was obtained using $\alpha_1 = 4$. This choice of hidden unit densities gives us a relative error below $3 \times 10^{-5}$ for the ground states and below $2 \times 10^{-4}$ for the excited states. It is necessary to compute the ground state accurately, since the error necessarily propagates to the excited state wavefunction due to the relation $\Psi = \Phi_{1} - \lambda \Phi_{0}$.  We also verified that the overlap with the ground state is below $1\%$ for a sample size of about $2000$.

We now turn to the Bose-Hubbard model in one-dimension with periodic boundary conditions, 
\begin{equation}
\hat{H} = -t \sum^{L}_{i=1} (\hat{c}^{\dagger}_{i}\hat{c}_{i+1} + \textrm{h.c.}) + \frac{U}{2} \sum^{L}_{i=1} \hat{n}_{i} (\hat{n}_{i} - 1),
\end{equation}
where $\hat{c}^{\dagger}_i$ and $\hat{c}_i$ are the boson creation and annihilation operators on site $i$, respectively, and $\hat{n} = \hat{c}_i^{\dagger}\hat{c}_i$ represents the local density at site $i$. For this problem, we experienced significant difficulty in lowering the relative error in both the 2-layer FFNN and the RBM even with a large number of hidden units, suggesting that either optimization is difficult or that the expressiveness of the ansatz is limited. A 3-layer FFNN, on the hand, converged significantly better.

We set $U=1$ and consider two system sizes. First the case of $N=10$  bosons on a one-dimensional periodic lattice with $L=10$ sites, for which exact results are easily obtained.  We used a 3-layer FFNN with hidden unit density $\alpha_1 = 4$ ($\alpha_2 = 1$) in the first (second) hidden layer ($860$ free parameters), and a RBM with hidden unit density $\alpha_1 = 8$ ($890$ free parameters). The relative error on the FFNN was lower than $5\times 10^{-4}$ for all momenta, whereas for the RBM one can see the error is increasing for larger momenta. The results are shown in Fig.~\ref{fig: Hubbard Spectrum}~(a).

Next, we show in Fig.~\ref{fig: Hubbard Spectrum}~(b) the results for $N=40$ bosons in $L=40$ sites. Here, the full (within the fixed particle number sector) Hilbert space dimension ($\sim 5 \times 10^{22}$) is too large to obtain results using ED. We could only infer the eigenenergies of the lowest few momentum sectors by matching with the lowest few eigenstates computed with MPS, since it is not straightforward to include momentum resolution in MPS.  Although MPS \cite{VerstraeteMPSsymmetry2013,VertraeteMPSSpin2018} can in principle be used to determine momentum spectra, it is challenging to efficiently use this method to higher dimensions and bosons.

\textit{Conclusions} --- We showed that artificial neural networks can be used as a variational Monte Carlo ansatz for obtaining excited states. In particular we showed two ways to achieve this: first, by using Abelian spatial symmetries such as translational symmetry and, second, by using a superposition of two neural networks such that the combined network represents a state orthogonal to the ground state. While the methods presented here were demonstrated using only simple networks (RBM and FFNN), they can in principle be used with any network architecture. More challenging models may require the use of more powerful networks, such as the convolutional arithmetic circuits or recurrent neural networks, which were shown to be highly efficient in representing highly entangled states~\cite{LevineShashua2018}. Our general strategy can also be generalized to represent other cases beyond spatial symmetries, for example permutational symmetry in fermionic systems would be a natural extension to pursue in future studies.

\section*{Acknowledgments}
KC was supported by the European Unions Horizon 2020 research and innovation program (ERC-StG-Neupert-757867-PARATOP). KC thanks the Flatiron Institute founded by the Simons foundation for hospitality.
The ED computations were carried out with the \textit{DiagHam} library.
The MPS computations were done using the ALPS package \cite{Dolfi2014}.
The ANN computations were based on NetKet \cite{NetKet}.

\bibliography{biblio}
\clearpage
\begin{widetext}
\section*{Supplementary Information: Network Optimisation Details}

\subsection{Stochastic reconfiguration}
Given a variational ansatz $\Psi(\lbrace \alpha_{k} \rbrace)$ we want to optimize the parameters $\alpha_{k}$ such that the trial wavefunction minimizes the energy of a target Hamiltonian $H$. There are several methods for achieving this task, the most common of which is stochastic gradient descent (SGD). We find however that an alternative method called stochastic reconfiguration (SR) has better performance especially when trying to find excited states. 

This optimization method was introduced by Sorella et al. in Ref.~\cite{Sorella2007SR} and can be seen as an imaginary time evolution. Let $\Psi(\lbrace \alpha^{0}_{k} \rbrace) \in \mathbb{C}^{2^n}$ be a wavefunction depending on an initial set of variational parameters $\lbrace \alpha^{0}_{k} \rbrace_{k=1, \dots, p}$. Consider now a small variation in the parameters $\alpha_{k} = \alpha^{0}_{k} + \delta \alpha_{k}$. The corresponding wavefunction can then be written as
\begin{equation}
\Psi(\lbrace \alpha_{k} \rbrace) = \Psi(\lbrace \alpha^{0}_{k} \rbrace) + \sum^{p}_{k} \delta \alpha_{k} \frac{\partial}{\partial \alpha_{k}} \Psi(\lbrace \alpha^{0}_{k} \rbrace).
\end{equation}
Introducing the logarithmic derivatives
\begin{equation}
\mathcal{O}_{k} = \frac{\partial}{\partial \alpha_{k}} \log\left[\Psi(\lbrace \alpha^{0}_{k} \rbrace)\right]
\end{equation}
the expansion can be rewritten as 
\begin{equation}
\Psi(\lbrace \alpha_{k} \rbrace) = \Psi(\lbrace \alpha_{k}^{0} \rbrace) + \sum^{p}_{k=1} \delta \alpha_{k} \mathcal{O}_{k} \Psi(\lbrace \alpha^{0}_{k} \rbrace).
\end{equation}
The  $\mathcal{O}_{k}$ are diagonal operators in the computational basis.

The SR scheme then proceeds by performing imaginary time evolution which to first order is given by
\begin{equation} \label{eqn: exp}
\Psi^{\prime}_{\mathrm{exact}} = (1 - \epsilon \hat{H}) \Psi.
\end{equation}
The aim now is to determine the coefficients $\lbrace \delta \alpha^{0}_{k} \rbrace_{k=1, \dots, p}$ corresponding to the new wavefunction $\Psi^{\prime}$, that minimizes the distance to $\Psi^{\prime}_{\mathrm{exact}}$ according to some chosen metric. For our simulations we used the Fubini-Study metric
\begin{equation}
\gamma(\phi, \Psi) = \arccos{\sqrt{\frac{\braket{\Psi|\phi} \braket{\phi|\Psi}}{\braket{\Psi|\Psi}\braket{\phi|\phi}}}}.
\end{equation}
After some algebra, we obtain to first order in $\epsilon$,
\begin{equation}\label{SR}
\sum_{k^{\prime}} \left[ \langle \mathcal{O}_{k}^{\dagger} \mathcal{O}_{k^{\prime}} \rangle -  \langle\mathcal{O}_{k}^{\dagger}\rangle \langle\mathcal{O}_{k^{\prime}}\rangle\right]\delta\alpha_{k^{\prime}} = -\epsilon \left[ \langle\mathcal{O}_{k}^{\dagger} \hat{H}\rangle -  \langle\mathcal{O}_{k}^{\dagger}\rangle \langle\hat{H}\rangle \right].
\end{equation}
This is the linear equation which we must solve for $\delta \alpha$. We then update the parameters as $\alpha_{k} = \alpha^{0}_{k} + \delta \alpha_{k}$ and repeat the procedure to convergence.

Since each SR iteration requires the inversion of a matrix, the computation complexity of each step is $\mathcal{O}(N_{w}^{3})$, as compared to $\mathcal{O}(N_{w})$ for the gradient descent methods, where $N_{w}$ is the number of variational parameters. However, the SR method is known to be more stable than the standard gradient descent methods. In preliminary studies, we have noticed that optimisation with SR requires much fewer iterations to converge. This faster convergence might make up for the larger computational cost required.

\subsection{Regularisation}
To perform each stochastic reconfiguration step, we need to solve Eq.~\eqref{SR}, which is basically a linear equation of the form
\begin{equation}
\boldsymbol{A} x = \boldsymbol{b}.
\end{equation}
The problem is that the matrices $\boldsymbol{A}$ and $\boldsymbol{b}$ are estimated using Monte Carlo sampling and thus is inherently noisy. Combined with the fact that the condition number of $A$ could be quite large, a small amount of noise could lead to a large error in $x$. This forces us to utilise some form of regularisation when solving the above inverse problem.

One possible method of regularisation is to add a multiple of identity to the matrix $\boldsymbol{A}$, i.e., $\boldsymbol{\widetilde{A}} = \boldsymbol{A} + \lambda \boldsymbol{I}$. We then solve for the system $\boldsymbol{\widetilde{A}} x = \boldsymbol{b}$, using the method of conjugate gradients. This is very similar to the well known Tikhonov or Ridge regularisation. For the simulations performed in the article, we typically choose $\lambda \in [0.001,0.01]$.

As a final remark, in the limit of large regularisation, $\lambda \gg \boldsymbol{A}$, the SR step Eq.~\eqref{SR} reduces to
\begin{equation}
\sum_{k'} \lambda \boldsymbol{I}_{kk'} \delta\alpha_{k'} =\lambda \delta\alpha_{k } = -\epsilon \left[ \langle\mathcal{O}_{k}^{\dagger} \hat{H}\rangle -  \langle\mathcal{O}_{k}^{\dagger}\rangle \langle\hat{H}\rangle \right]
\end{equation}
which simply corresponds to the standard stochastic gradient descent with a learning rate $\frac{\epsilon}{\lambda}$.

\subsection{Monte Carlo Sampling}
In this work, the Monte Carlo samples are obtained using the standard Metropolis-Hastings algorithm \cite{Metropolis1953} which is a Markov chain Monte Carlo method. Each iteration of the algorithm proceeds as follows:
\begin{enumerate}
  \item Initialisation: We begin with a random configuration $\boldsymbol{\sigma}_{1}$.
  \item At each iteration $t$: We propose a new configuration $\boldsymbol{\sigma'}$.
  \item Accept/Reject: By evaluating the network, compute the ratio $p = \left| \textrm{exp}[\log(\Psi(\boldsymbol{\sigma'})) - \log(\Psi(\boldsymbol{\sigma_{t}}))] \right|^{2} =\left| \frac{\Psi(\boldsymbol{\sigma'})} {\Psi(\boldsymbol{\sigma}_{t})} \right|^{2}$
  \begin{itemize}
  \item Accept the configuration $\boldsymbol{\sigma'}$ with probability $p$ and reject with probability $1-p$
  \item If we accept: $\boldsymbol{\sigma}_{t+1} = \boldsymbol{\sigma'}$
  \item If we reject: $\boldsymbol{\sigma}_{t+1} = \boldsymbol{\sigma}_{t}$
  \end{itemize}
\end{enumerate}
In this way, we generate a chain of configurations $\lbrace \boldsymbol{\sigma}_{1}, \boldsymbol{\sigma}_{2}, \boldsymbol{\sigma}_{3}, \cdots \rbrace$ which in the infinite limit would correspond to a sample drawn from our target distribution 
\begin{equation}
P(\boldsymbol{\sigma}) = \frac{|\Psi(\boldsymbol{\sigma})|^{2}}{\sum_{\boldsymbol{\sigma}} |\Psi(\boldsymbol{\sigma})|^{2}} \propto e^{2\textrm{Re}[\log(\Psi(\boldsymbol{\sigma}))]}.
\end{equation} 
After obtaining such a Markov chain, we can pick every $n^{\textrm{th}}$ configuration (where $n$ is ideally larger than than the correlation time of the chain) to form our sample. For our simulations, we use at least as many samples as there are free parameters in the network, typically between $1000$ to $10000$.

Because the wavefunctions for excited states are generally not smooth, we implemented the parallel tempering method \cite{ParallelTempering1986} first introduced by Swendsen and Wang, in order to obtain a more accurate sampling. Here, we essentially run $N$ copies of the above method,  i.e. $N$ Markov chains, but at $N$ different temperatures. That is, in step 3, we replace $ p \rightarrow \left| \textrm{exp}(\beta[\log(\Psi(\boldsymbol{\sigma'})) - \log(\Psi(\boldsymbol{\sigma_{t}}))])\right|^{2}$ which would give us a sample drawn from the distribution 
\begin{equation}
P(\boldsymbol{\sigma})  \propto e^{2\beta\textrm{Re}[\log(\Psi(\boldsymbol{\sigma}))]}
\end{equation}
where $\beta$ can be interpreted as inverse temperature, with $\beta=1$ corresponding to our target distribution. Now, in addition to the moving along the $N$ chains according to the 3 steps described above, we have to consider exchanges between the chains. The full algorithm for each iteration is then
\begin{enumerate}
\item Perform the above Metropolis-Hasting iteration $m$ times for each of the $N$ Markov chains where chain $i$ has temperature $\beta_{i} = \frac{N-i}{N}$, such that chain $0$ corresponds to our target temperature.
\item Starting from even indices $i$, we exchange the current configuration of chain $i$ and $i+1$ with a probability
\begin{equation}
p = e^{(\beta_{i+1} - \beta_{i})\left\{2\textrm{Re}[\log(\Psi(\boldsymbol{\sigma_{i}})) - \log(\Psi(\boldsymbol{\sigma_{i+1}}))]\right\}}
\end{equation}
\item Do the same for odd indices $i$.
\end{enumerate} 
For the simulations in the article, where parallel tempering is necessary, we typically use between $100$ to $200$ replicas. We did notice the following caveat: If we just use the standard Metropolis-Hastings algorithm, it can happen the Monte Carlo sampled energy differ significantly from the one obtained using parallel tempering or where possible the exactly computed energy (by summing over all configurations). Still, parallel tempering with sufficient number of replicas has always managed to obtain an energy close to the exactly computed one. 

A final point to note is that the choice of the temperature set has an important role in the efficiency of the sampling process. In our implementation, we simply chose equally spaced temperature points which is known to be far from optimal. Methods have been introduced to adaptively and iteratively optimise the temperature set \cite{OptimisedPT2006}. 

\subsection{Summary of Optimisation Parameters}
For the simulations done in this paper, the main optimisation parameters are: (1) regularisation parameter $\lambda$, (2) learning rate $\epsilon$, (3) Number of replicas for parallel tempering, (4) Number of samples for the Monte Carlo average, (5) standard deviation, $\sigma$, of initial parameters (the parameters are initialised with a normal distribution with a zero mean value and a standard deviation of $\sigma$).
The choice of these parameters for the various simulations are given in the table below.

\begin{table}[h]
\begin{tabular}{|l|l|l|l|l}
\cline{1-4}
                         & RBM           & RBM (Gap)         & 3-layer FFNN         &  \\ \cline{1-4}
Regularisation, $\lambda$ & 0.01          & 0.001             & 0.01         &  \\ \cline{1-4}
Learning Rate, $\epsilon$            & 0.01          & 0.01              & 0.01         &  \\ \cline{1-4}
Number of Parallel Tempering Replicas           & n.a.          & n.a.              & 100-200         &  \\ \cline{1-4}
Number of Samples        & \multicolumn{3}{l|}{$\sim$ number of parameters} &  \\ \cline{1-4}
Initial Standard Deviation, $\sigma$           & 0.01          & 0.01              & Layer $n$: $0.3/M_{n}$        &  \\ \cline{1-4}
\end{tabular}
\end{table}

\subsection{Computational Time: ANN vs ED}
In Fig.~\ref{fig: time}, we compare the computational time of the RBM versus that of ED. For the system sizes presented here, the RBM generally requires more computational time and resource. However, it is clear that while the time for the ED calculations scales exponentially with system size, the RBM only scales polynomially. The FFNN with a similar number of parameters as the RBM generally takes a similar amount of time. 
\begin{figure}[h]
            \includegraphics[width=0.4\textwidth]{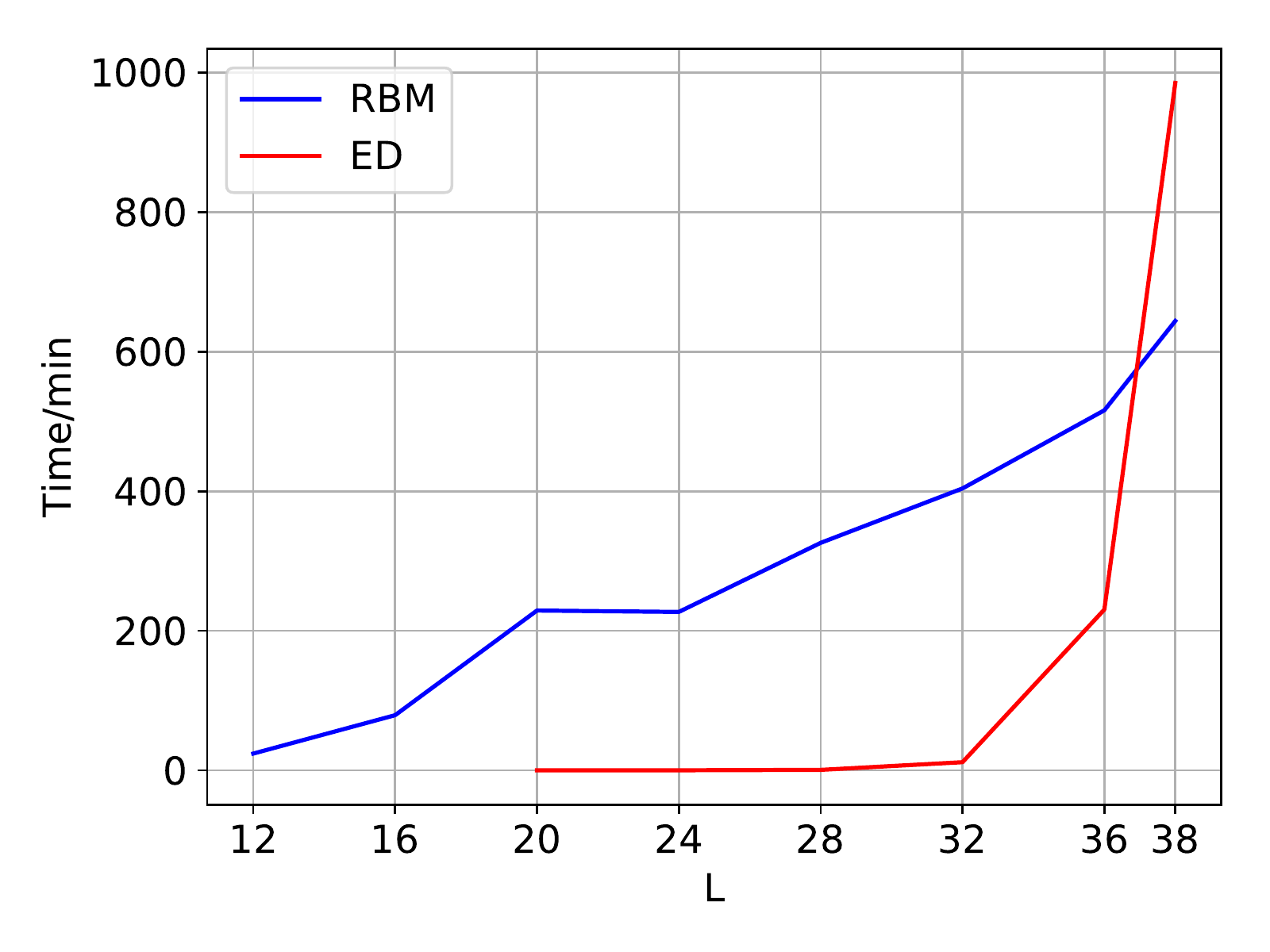}
    \caption{Computational time of the RBM and ED for the one-dimensional spin half Heisenberg model. The RBM used has a hidden unit density of $\alpha_1 = 3$. The number of samples used is fixed  at $5000$ and the number of iterations is $10000$. The computational time shown for RBM and ED is the wall time of the whole process executed on 28 cores of the Intel(R) Xeon(R) CPU E5-2690 v4 and 12 cores of the Intel Dual Xeon 2630 respectively.}
 \label{fig: time}
\end{figure}

\subsection{Convergence}
In Fig.~\ref{fig: samples} show here the typical convergence properties of the two schemes introduced in the main text.
\begin{figure}[h]
            \includegraphics[width=0.8\textwidth]{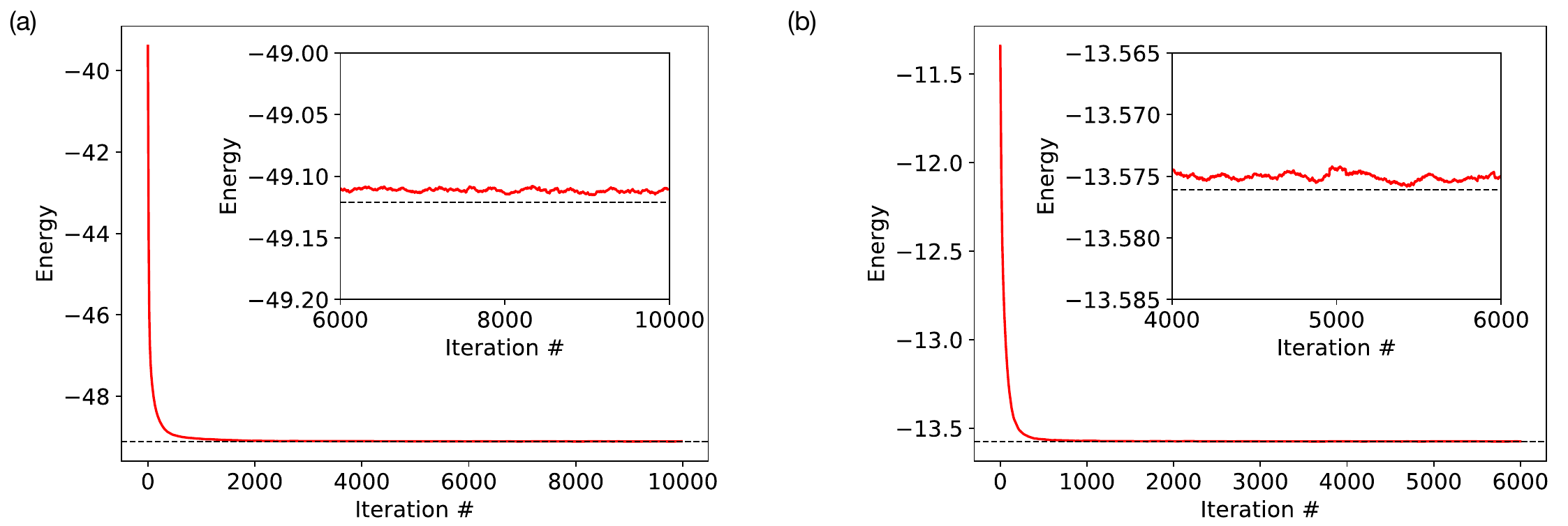}
    \caption{Convergence properties of the optimisation scheme. The red line shows the average energy of the previous 100 iterations. Dash lines indicates the exact values. (a) First excited state of the $L=28$ one-dimensional spin half Heisenberg model using the 2-step method involving the sum of two RBM. (b) Bose Hubbard model with $N=10$ bosons in $L=10$ sites at momentum $k =8\pi /10$ using the two-layer FFNN. }
 \label{fig: samples}
\end{figure}

\end{widetext}

\clearpage
\end{document}